# On the significance of the excesses in the ATLAS diphoton and four lepton decay channels

Gioacchino Ranucci

*Istituto Nazionale di Fisica Nucleare*
*Via Celoria 16 - 20133 Milano*
*Italy*
Phone: +39-02-50317362
Fax: +39-02-50317617
e-mail: [gioacchino.ranucci@mi.infn.it](gioacchino.ranucci@mi.infn.it)

**Abstract**

This note describes an assessment of the statistical significance of the recently released ATLAS data regarding the Higgs search in the decay channels especially suited for the low mass region, in particular the diphoton and four lepton decay channels. Besides providing the significance evaluation of the considered channels, alone or combined, in the low mass range from 110 to 146 GeV, some details of the profile likelihood procedure employed for the calculations are described, too. Special emphasis is given to the *look elsewhere effect*, which arises when the search of a new signal is carried out over a broad mass range, therefore specifying separately *local* and *global* significances. When combined together, the global *p-value* of the localized excesses detected in the two channels is found equal to either 0.033 or 0.013, corresponding, respectively, to a significance of $1.8\sigma$ and $2.2\sigma$, depending upon the assumption of mutual independence or dependence of the amplitudes of the expected signal in each channel. Among the other obtained results, the local and global significances of the two channels individually taken are computed in good agreement with those reported by the Collaboration, while only a minor discrepancy is found for their combined local significance. Finally, some considerations are illustrated related to the low statistics four lepton decay channel, showing that a search strategy different from the profile likelihood method, e.g. scan statistics, can result in a substantially different significance, enhanced to a $3.3\sigma$ global effect.

The ATLAS experiment has recently released the results on the Higgs search based on the data accumulated so far in three channels featuring sensitivity in the low mass range [1], [2] and [3]:

$H \to \gamma\gamma$ , $H \to ZZ^{(*)} \to \ell^+\ell^-\ell^+\ell^-$ and $H \to WW^{(*)} \to \ell^+\nu\ell^-\bar{\nu}$

While the search in the third channel is characterized by a broad excess of events over the background expectation, the other two channels exhibit localized excesses of local significance evaluated by the ATLAS Collaboration as high as $2.8\sigma$ and $2.1\sigma$, respectively, which remarkably happen to be essentially coincident in mass. The obvious question raised by this peculiar data occurrence is the following: which is the probability that a configuration as extreme, or even more extreme, than that actually observed may be originated by an unlucky background fluctuation? By definition, such a probability is the *p-value* of the observed configuration, which following the common (but not indispensable) practice is then converted into a significance level by the usual association with the one sided tail probabilities of the Normal distribution.

The evaluation of the desired *p-value* must be performed taking into account the now famous *look elsewhere effect* [4], e.g. it is not enough to evaluate the local probability of the observed excess, but instead one has to compute the probability for that excess to appear anywhere in the examined mass range.

A coherent framework to determine local and global *p-values* (and significances) in the search of new signals is provided by the profile likelihood ratio algorithm [5].

In essence this method relies on the definition of the profile likelihood ratio as test statistic, i.e. as criterion to discern whether there is a signal embedded in a continuous background distribution. While referring the interested readers to the abundant literature on the subject, succinctly it may be reminded that the considered test statistic (from now on indicated as LR) is actually the logarithm (multiplied by -2) of the ratio of the likelihood of the observed data configuration under the two alternative hypotheses of background-only and of signal+plus background; when LR assumes values close to zero, it indicates absence of signal, and vice versa it takes positive values much higher than zero when a signal is present. Between these two extreme cases the LR test statistic features a continuous distribution (the sampling distribution in statistical language), which is the basis to quantitative determine the *p-value* of any observed data configuration.

How this is practically accomplished is explained through the concrete application to the ATLAS diphoton distribution, whose characteristics are derived from the plots and associated information in [2]. Differently from the approach followed by the Collaboration, which constructs the combined likelihood function from the likelihood functions of the nine categories in which the data are divided, here a simpler procedure is adopted to write directly the likelihood of the global diphoton distribution (figure 1 of reference [2]).

The method relies on a limited set of simulations performed through a simplified MC. In each simulation a diphoton mass distribution is generated following the exponential background behavior found by the ATLAS collaboration, and with no Higgs signal embedded in it. Then the simulated data set is subjected to the same search procedure that later on one would like to apply to the real data, e.g. it undergoes a continuous scan, through the profile likelihood method, over the mass range of interest to look for the possible presence of small excesses of events. Concretely this means that the likelihood ratio is computed locating the hypothesized signal at each mass, in practice a fine grid, of the explored interval. In this search the amplitude of the putative signal can float freely, being effectively a fit parameter. Taking into account the results of the recent LHC combined Higgs boson search [6], the adopted search range is restricted to the 110-146 GeV interval not excluded at the 99% confidence level. The width of the hypothetical signal is taken from the MC estimates reported in [2].

To illustrate the approach, Fig. 1 displays one of these simulations (1a) and the corresponding behavior of the LR test statistic across the spanned energies (Fig. 1b). It can be appreciated that where there are more positive wiggles in the simulated data, correspondingly the LR quantity assumes a value remarkably different from zero. Since the criterion to discern the presence of a signal is specifically based on the ordinate of the highest peak of LR, we need to know how extreme such a peak can be induced by the background alone. This information can be obtained by repeating many (time consuming) simulations, but there are methodologies that allow to determine the sampling distribution (entirely or the tail only) of the highest peak of LR with the help of a reduced set of simulations [7][8]. By using, in particular, the method described in [8], the LR sampling distribution estimated on the basis of 1000 MC iterations is shown in Fig. 2, together with the MC output. The long tail is the subtle characteristic of this curve, implying that there are some background occurrences that can originate a very high value of LR, and thus mimicking a true signal.

Now let's consider the actual diphoton ATLAS data: the measured distribution and the corresponding behavior of the test statistic LR are shown together in Fig. 3a and 3b.

In particular the peak value of the test statistic in Fig. 3b, found at 126 GeV, is 7.5. By integrating the sampling distribution reported in Fig. 2 above the 7.5 abscissa one obtains 0.068, i.e.

in 6.8% of the cases a pure background distribution would originate a LR test statistic peak value higher than that inferred from the actual data. This means that the excess observed at 126 GeV in the diphoton mass distribution *per se* is not so peculiar, but is well in the realm of the possible background outcomes. The *p-value* 0.068 corresponds to a significance of 1.5σ, which, in summary, is the global significance over the explored mass range (110-146 GeV) of the observed event excess in the diphoton channel.

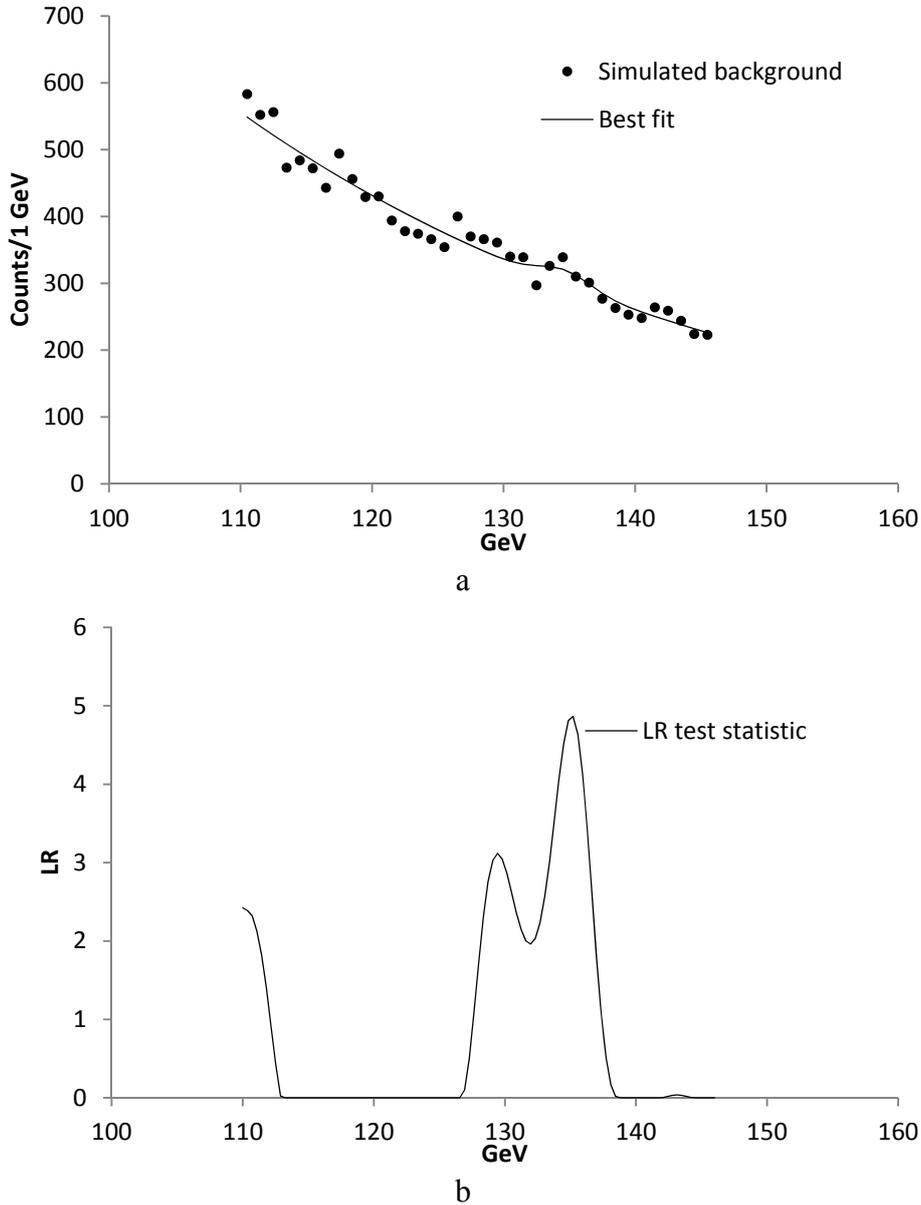

*Fig. 1 – a) Example of a background-only dataset simulated according to the diphoton distribution characteristics and b) the corresponding plot of the LR test statistic.*

For completeness, and as consistency check, the same likelihood ratio procedure has been exploited to assess also the local significance: this has been done repeating the same Monte Carlo calculation, but at a fixed presumed mass value for the signal. The value of the LR test statistic is found higher than the actual experimental value, 7.5, in a 0.27% fraction of the total simulations performed. This is, thus, the local *p-value* of the excess, corresponding approximately to a local significance of 2.8 σ, very close to the estimate of 2.74σ obtained by the simple square root of the 7.5 peak value, on the basis of the prescription stemming from the asymptotic formulae for the test statistic [5].

Both the local and global significances inferred from the simplified procedure followed here are in pretty good agreement with the values reported by the Collaboration in [2].

It should be pointed out that the calculation performed as described is based on the standard approach to impose that the amplitude of the searched fluctuation over the background is positive, as it should be if there were a true signal embedded in the data. However, since the background alone can fluctuate both upward and downward, the full statistical deviation from its expected value can also be quantified removing the restriction that the amplitude parameter must be positive. By repeating the previous procedure under this condition, the resulting global *p-value* is 0.105, i.e. the probability of a positive/negative background–induced fluctuation anywhere in the examined range, as extreme or more extreme in absolute value than that actually observed, is 10.5%.

We then consider the four lepton channel, again with the restriction to the interval 110-146 GeV. The distinctive feature of this low statistic channel is that in the mass range of interest only three events occur, but clustered within 1GeV; their respective energy is 123.6, 124.3 and 124.6 GeV. The data are considered binned in bins of 1 GeV width, therefore there is one count in the bin 123-124 and there are two counts in the bin 124-125. Background and signal resolution information are derived from the plots and data in [1].

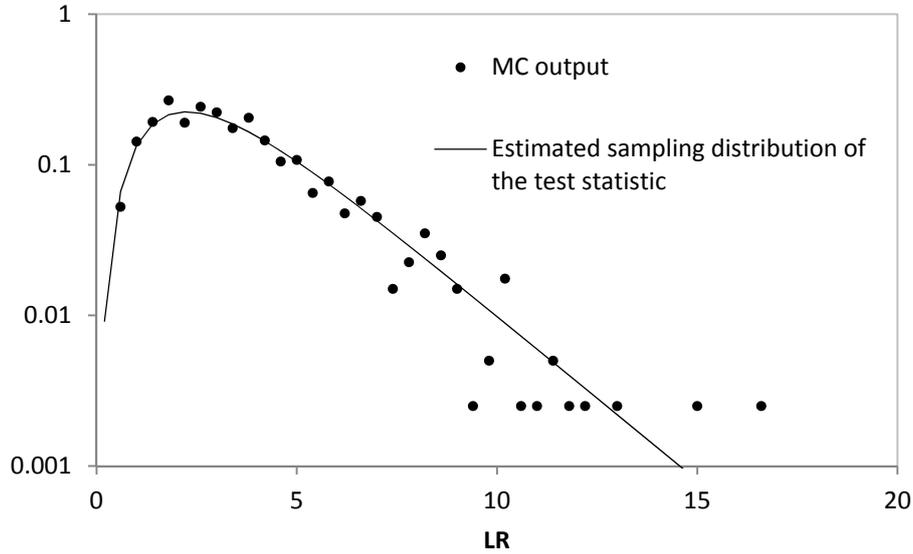

*Fig. 2 – Estimated sampling distribution of the test statistic LR for the diphoton ATLAS data obtained via a limited set of 1000 Monte Carlo iterations.*

By applying the same profile likelihood procedure employed for the diphoton channel (for the sake of brevity the details are omitted) the global *p-value* over the considered mass range is found 21%, while the local *p-value* is evaluated to be about 0.024 (significance 2σ), to be compared with the values, respectively, O(30%) and 2.1σ reported by the Collaboration in [1].

Finally, the global significance of the excesses in the two channels considered together can be obtained exploiting the joint profile likelihood, i.e. a unique likelihood constructed for both channels. In order to keep a general approach, and taking into account the very low statistics of the four lepton channel, in a first instance the amplitude parameters for the signal on the two channels are left independent, thus not mutually constrained by the relevant Standard Model Higgs branching ratios. The plot of the joint LR is shown in fig 4a, featuring a peak value of ordinate 10.3 at 125.5 GeV.

Through a Monte Carlo procedure like that used previously, with 1000 iterations an estimate of the sampling distribution of the joint test statistic is obtained, and is shown in fig 4b together with the simulation result. The integral of such a distribution above the abscissa 10.3, i.e. the peak value of the LR plot in fig 4a, is 0.033, which is thus the global *p-value* of the coincident configuration observed on the two channels, corresponding to 1.8σ.

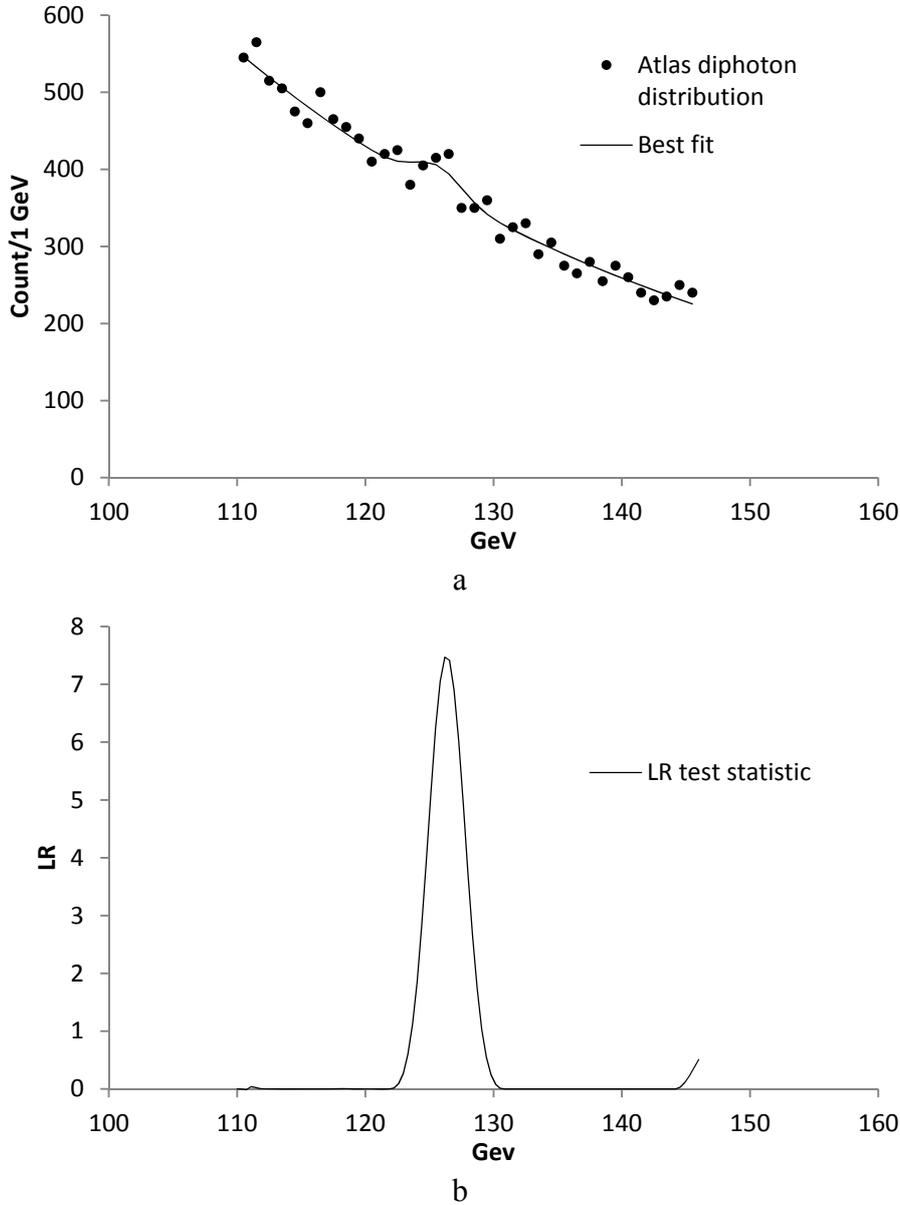

*Fig. 3 – a) ATLAS diphoton data displayed together with the best fit curve and b) the corresponding plot of the LR test statistic.*

Therefore, according the profile likelihood method, a concurrent occurrence caused anywhere in the examined range by the background, as extreme or more extreme than that actually detected jointly in the diphoton and four lepton channels, is not a very rare outcome, but it may happen with a 3.3% probability.

Similarly, also the joint local significance can be assessed; through a MC in which the hypothesized signal is kept at a fixed mass, the joint local *p-value* of the diphoton and four lepton channels is determined to be 0.0018, thus slightly enhanced with respect to the 0.0027 *p-value* featured by the diphoton channel alone, and obviously more enhanced when compared to the individual 0.024 *p-value* of the 4 lepton channel. The *p-value* of 0.0018 corresponds to a significance of 2.9σ, showing thus a discrepancy with the 3.4σ local significance for these two channels together reported in [9], mostly due to the different treatment of the relationship between the amplitude parameters of the two channels.

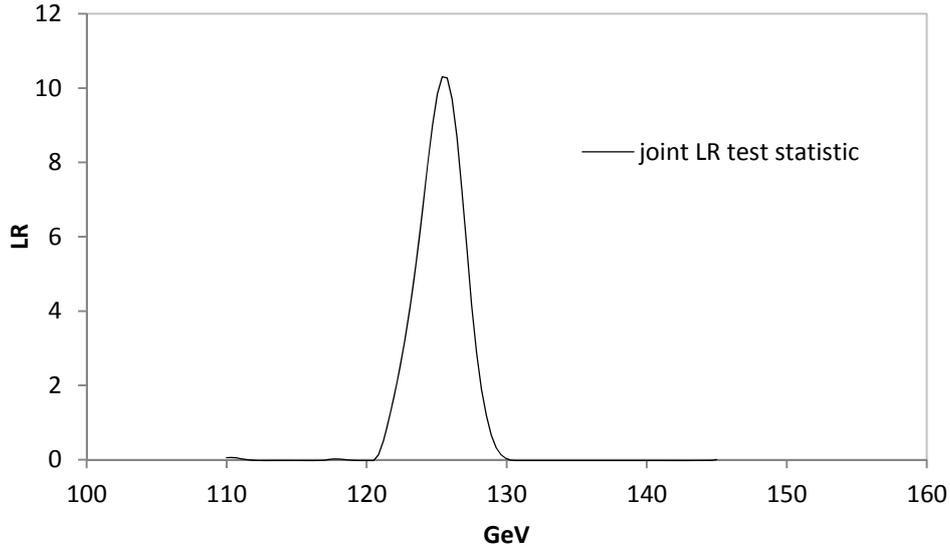

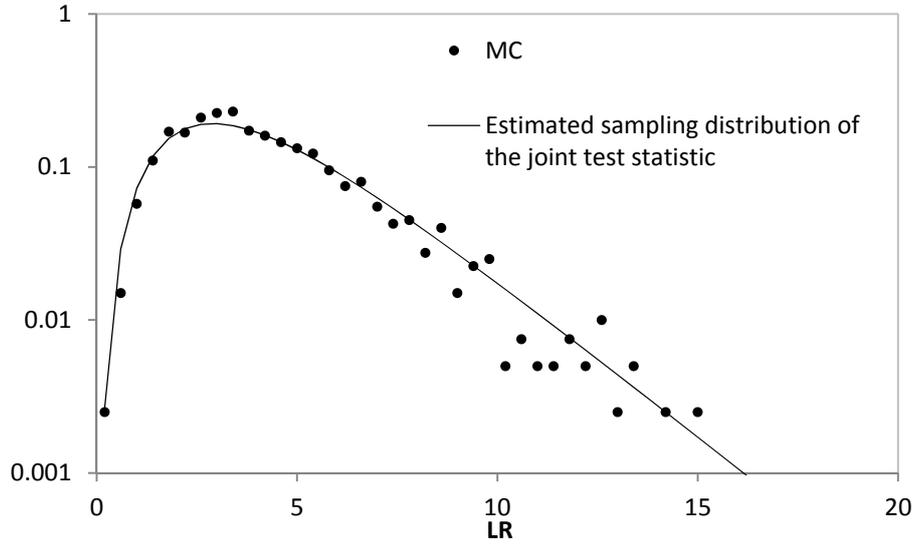

*Fig. 4 – a) Plot of the joint LR test statistic of the ATLAS diphoton and four lepton channels and b) the corresponding estimated sampling distribution obtained via a set of 1000 Monte Carlo iterations.*

Indeed, if following the standard method [10] the two parameters are considered linked according to the respective expected branching ratios (this amounts to have effectively only one amplitude parameter in the joint likelihood), by repeating the entire calculation the joint local *p-value* is found enhanced to 0.0008, corresponding to a significance of about 3.2σ.

In order to cross check the correctness of this evaluation, it has been verified by MC that the *local* (i.e. at fixed mass) sampling distribution of the test statistic stemming from the combined likelihood of the two channels, written with a unique amplitude parameter, follows the expected asymptotic distribution. That this is actually the case can be appreciated from Fig. 5, which displays together the expected $\chi^2(1)$ distribution and the MC output of the test statistic for all the simulated occurrences (about half of the total) in which the amplitude parameter from the fit is found greater than 0: the agreement is really excellent.

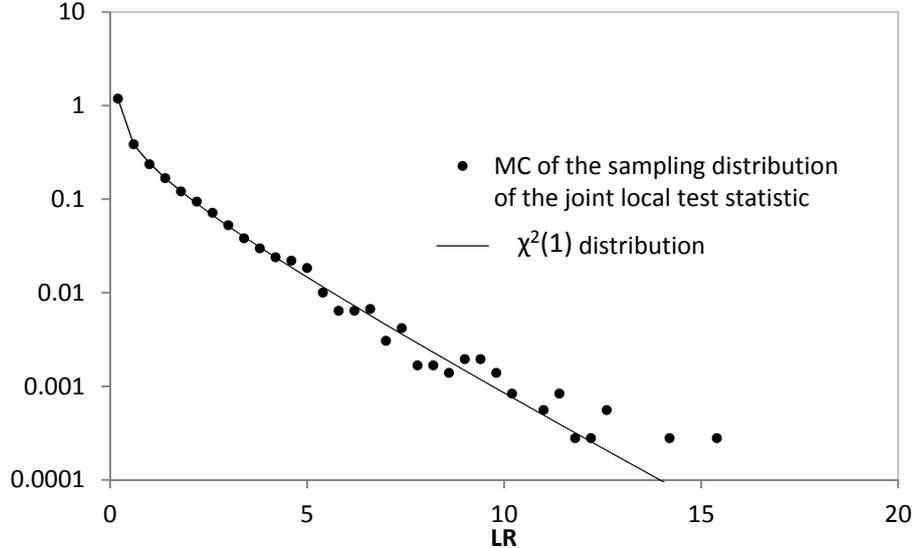

*Fig.5 – Monte Carlo evaluation and expected asymptotic distribution of the joint local test statistic, in the case of amplitude parameters in the two channels linked via the Standard Model branching ratios. The agreement between them is excellent.*

For completeness, also the significance with the look elsewhere effect included is re-computed under the same assumption of a common amplitude parameter; the global *p-value* determined in this way is 0.013 (2.2σ), hence again more significant with respect to the previous calculation performed under the scenario of two independent parameters.

Since the evaluation carried out by the ATLAS Collaboration is done in the framework of a unique amplitude parameter, the last result can be compared with that reported by the Collaboration in [9] for the joint global significance of all the three channels listed at the beginning of this note (the combination of the diphoton and four lepton channels only is not available). The result pinpointed by the Collaboration after the look elsewhere effect corresponds to a *p-value* of about 0.6%, which appears reasonably consistent with the 1.3% determined in this work, taking into account that the impact of the third channel is expected to be modest.

Before completing this discussion, some additional considerations can be done on the low statistics four lepton channel. First of all, why a so tight occurrence of three events within 1 GeV is judged by the profile likelihood method background-compatible with a probability as high as 21% (or even of the order of 30%, as found by the Collaboration)? The fact is that the profile likelihood compares the no signal case with a hypothetical signal configuration, in which the counts may be spread over a sizable interval as effect of the resolution, implying that the extreme tightness in mass of the three detected events does not play a very special discriminating role in the algorithm.

An alternative assessment which weighs more the contiguity of the three events, instead, can be obtained by examining the possible configurations of the background alone, with no reference to any signal hypothesis. In particular, adhering to the usual definition of *p-value* as the probability of all the background-induced configurations as extreme or more extreme than the one detected, an approximate determination of its value in this different framework can be inferred through the calculation of the probability that in any of the 36 bins from 110 to 146 the background originates 3 or more than 3 counts. In this respect, for a simple order of magnitude evaluation avoiding an unnecessary (at this stage) calculation burden, the four lepton data are still considered binned in bins of size 1 GeV, but the background is assumed flat, amounting to the level of 0.12 counts/GeV obtained averaging the values from 110 to 146 GeV read from the figure 4a of [1].

With this simplification, the probability function for the maximum number of detected counts $N$ across the considered bins is given by the following formula [11]

$$P_{\max}(N) = \sum_{k=1}^{W} \binom{W}{k} \left( \sum_{n=0}^{N-1} \frac{e^{-B} B^n}{n!} \right)^{W-k} \left( \frac{e^{-B} B^N}{N!} \right)^{k}$$

where $W$ is the number of bins and $B$ the average background in each bin.

Fig. 6 displays the Poisson probability in a single bin and the probability of the maxim count detected over all the 36 bins, as stemming from the formula. By summing in the latter all the probabilities for the counts greater and equal than 3, one would obtain as global *p-value* 0.0094, definitively lower than the result obtained with the profile likelihood. By the way, the rule of thumb to determine the global *p-value* by taking the *p-value* for single bin and multiply by 36, valid when the highest count is much higher than the mean value, is recovered by the previous formulation.

Anyhow, this treatment is an approximation of the correct approach based on the use of scan statistics. Scan statistics avoids any form of quantization (binning) of the problem: it simply assumes a sliding window exploring the range of interest, affected by a smooth background, and provides an answer to the question of the probability of the maximum cluster of background-induced events found in the observation window along the scanning process.

This procedure is more precise than the fixed bin approach, since it takes into account configurations otherwise missed, like 2 counts in a bin and 1 count in the adjacent bin, but with distance below the bin size, therefore the *p-values* obtained with the above fixed bin method are to be considered a lower bound for the true *p-values*.

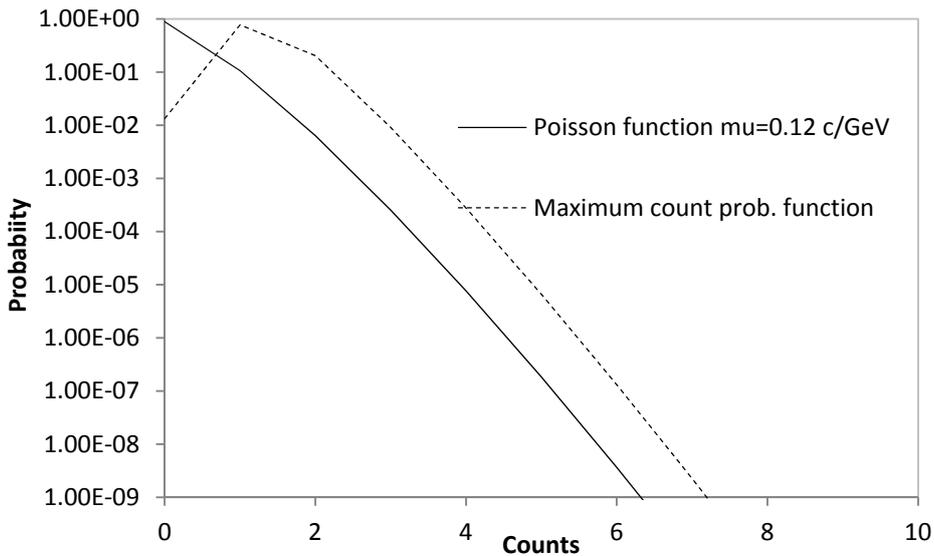

*Fig. 6 – Plots of the single bin Poisson probability function and of the resulting probability function of the maximum detected count across the 36 bins in which the 100-146 GeV range is divided.*

Interested readers can refer to [12][13] and the literature indicated there for a discussion of scan statistics. In particular, several approximated formulas are available to compute the desired *p-values* associated with the sliding window procedure; for the present evaluation the Naus approximation [14] has been used, obtaining a *p-value* equal to 0.027 (about $1.9\sigma$ global significance), confirming that the fixed bin approach is actually originating in this respect a lower limit estimate.

In order to understand the implications of the simplified assumption on the background, a conservative upper bound can be estimated adopting as uniform rate the maximum value of the background in the 110-146 GeV interval, 0.14 counts/GeV, instead of the average value; the resulting upper limit for the *p-value* is equal to 0.039. This evaluation indicates that the more precise determination that could be done considering the actual background shape would not differ much from that stemming from the averaged background approximation.

It can, thus, be concluded that 0.027 is a good approximation of the global *p-value* for the four lepton channel in the framework of the sliding window exploration process, expressing directly the probability to obtain anywhere in the examined range, as effect of the background, the kind of excess actually observed in the data, i.e. three events within a 1 GeV interval. It is remarkably lower than the *p-value* obtained with the profile likelihood approach, weighting more this channel in term of significance with respect to the other two.

This argument in principle can be further pursued by considering the two channels as elements of a coincident measurement scheme in which, upon detecting an effect in the diphoton channel, an observation interval of 6 GeV width is open in the four lepton channel, centered around the mass of the effect detected in the first channel. Since it is desired to evaluate the probability of the background configurations as extreme or more extreme than that actually occurred, the condition imposed on the coincidence interval is not simply to contain three events, but three events clustered within 1 GeV. The scan statistics approach quantifies this probability as 0.61%. By reminding that the probability of a chance fluctuation in the first channel at or above the observed level is 6.8%, the global probability of the joint configuration computed in this way is 0.04% (3.3σ). Therefore, the observed data configuration is judged less background-compatible if compared with the evaluation based on the profile likelihood methodology.

In conclusion, the significance of the excess detected in the recently released diphoton and four lepton channels ATLAS data has been computed following the profile likelihood method in the low mass range from 110 to 146 GeV, obtaining a global *p-value* after the look elsewhere effect of 1.3% (2.2σ), in the context of the standard analysis scheme in which the signal amplitude parameters on the two channels are linked via the Standard Model branching ratios. By adopting the alternative scan statistics search methodology on the four lepton channel, the *p-value* would be reduced to 0.04%, promoting the joint excess to a 3.3σ global effect.

**Acknowledgements**
I would like to thank M. Fanti for extremely useful discussions about the subject of this manuscript.

**References**

[1] ATLAS Collaboration, *Search for the Standard Model Higgs boson in the decay channel $H \rightarrow ZZ^{(*)} \rightarrow \ell\ell\ell\ell$ with 4.8 $fb^{-1}$ of pp collisions at $\sqrt{s} = 7$ TeV*, ATLAS-CONF-2011-162 (2011)

[2] ATLAS Collaboration, *Search for the Standard Model Higgs boson in the diphoton decay channel with 4.9 $fb^{-1}$ of ATLAS data at $\sqrt{s} = 7$ TeV*, ATLAS-CONF-2011-161 (2011)

[3] ATLAS Collaboration, *Search for the Higgs boson in the $H \rightarrow WW^{(*)} \rightarrow \ell^+\nu\ell^-\bar{\nu}$ decay channel in 2.05 $fb^{-1}$ of pp collisions at $\sqrt{s} = 7$ TeV with the ATLAS detector*, submitted to Phys. Rev. Lett. (2011)

[4] L. Demortier, *P Values and Nuisance Parameters*, Proceedings of PHYSTAT-LHCWorkshop, 2007, CERN-2008-001

[5] G. Cowan, K. Cranmer, E. Gross and O. Vitells, *Asymptotic formulae for likelihood-based tests of new physics*, Eur. Phys. J. **C71** (2011) 1–19

[6] ATLAS and CMS Collaborations, *Combined Standard Model Higgs boson searches with up to 2.3 $fb^{-1}$ of pp collisions at $\sqrt{s}=7$ TeV at the LHC*, ATLAS-CONF-2011-157, CMS-PAS-HIG-11-023 (2011)

[7] E. Gross and O. Vitells, *Trial factors for the look elsewhere effect in high energy physics*, Eur. Phys. J C **70** (2010) 525–530

[8] G. Ranucci, *The profile likelihood ratio and the look elsewhere effect in high energy physics,* Nuclear Instruments and Methods in Physics Research A, Volume 661, Issue 1, p. 77-85 (2012) , e-print: arXiv:1201.4604

[9] ATLAS Collaboration, *Combination of Higgs Boson Searches with up to 4.9 $fb^{-1}$ of pp Collision Data Taken at $\sqrt{s}$ = 7 TeV with the ATLAS Experiment at the LHC*, ATLAS-CONF-2011-163 (2011)

[10] ATLAS Collaboration, Expected performance of the ATLAS experiment, detector, trigger and physics. CERN-OPEN-2008-020, Geneva (2008). e-print: arXiv:0901.0512

[11] G. Ranucci, *An alternative view of the look elsewhere effect*, in Proceedings of the PHYSTAT 2011 Workshop on Statistical Issues Related to Discovery Claims in Search Experiments and Unfolding, CERN, Geneva, Switzerland, 17–20 January 2011, edited by H.B. Prosper and L. Lyons, CERN–2011–006, pp. 190–198

[12] F. Terranova, *Peak finding through Scan Statistics*, Nuclear Instruments and Methods in Physics Research Section A, Volume 519, Issue 3, p. 659-666 (2004)

[13] K. J. Orford, *The analysis of cosmic ray data*, J. Phys. G: Nucl. Part. Phys. 26 (2000) R1–R26

[14] J.I. Naus, *Approximations for distributions of scan statistics*, J. Amer. Stat. Ass. 77 (1982) 177